\documentclass[11pt,a4paper,pdftex]{article}
\pdfoutput=1
\usepackage{jcappub}
\usepackage{color}

\newcommand{\be}{\begin{eqnarray}}
\newcommand{\ee}{\end{eqnarray}}
\newcommand{\rar}{\rightarrow}

\title{Unattainable extended spacetime regions in conformal gravity}

\author[a]{Hrishikesh~Chakrabarty,}
\author[a]{Carlos~A.~Benavides-Gallego,}
\author[a,b,1]{Cosimo~Bambi%
\note{Corresponding author}}
\author[c]{and Leonardo~Modesto}

\affiliation[a]{Center for Field Theory and Particle Physics and Department of Physics,\\
Fudan University, 2005 Songhu Road, 200438 Shanghai, China}
\affiliation[b]{Theoretical Astrophysics, Eberhard-Karls Universit\"at T\"ubingen,\\ 
Auf der Morgenstelle 10, 72076 T\"ubingen, Germany}
\affiliation[c]{Department of Physics, Southern University of Science and Technology (SUSTech),\\ 1088 Xueyuan Road, Shenzhen 518055, China}

\emailAdd{17110190059@fudan.edu.cn}
\emailAdd{17110190058@fudan.edu.cn}
\emailAdd{bambi@fudan.edu.cn}
\emailAdd{lmodesto@sustc.edu.cn}

\abstract{The Janis-Newman-Winicour metric is a solution of Einstein's gravity minimally coupled to a real massless scalar field. The $\gamma$-metric is instead a vacuum solution of Einstein's gravity. Both spacetimes have no horizon and possess a naked singularity at a finite value of the radial coordinate, where curvature invariants diverge and the spacetimes are geodetically incomplete. In this paper, we reconsider these solutions in the framework of conformal gravity and we show that it is possible to solve the spacetime singularities with a suitable choice of the conformal factor. Now curvature invariants remain finite over the whole spacetime. Massive particles never reach the previous singular surface and massless particles can never do it with a finite value of their affine parameter. Our results support the conjecture according to which conformal gravity can fix the singularity problem that plagues Einstein's gravity.}


\begin{document}

\maketitle


\section{Introduction}

Einstein's gravity was proposed more than 100~years ago and is still the current framework for the description of the gravitational field and of the chrono-geometrical structure of the spacetime. The theory has passed a large number of observational tests, and current data agree well with theoretical predictions~\cite{tests1,tests2}. Nevertheless, there are a few important problems demanding new physics. One of these open issues is that physically relevant solutions in Einstein's gravity have spacetime singularities, where predictability is lost and standard physics breaks down. While there have been many efforts to figure out how to solve the singularity problem, no attempt has been completely successful so far.

Among the proposals to solve the singularity problem in Einstein's gravity, here we are interested in the family of conformally invariant theories of gravity~\cite{confgrav1,confgrav2,confgrav3,confgrav4,confgrav5,confgrav8,confgrav9}. 
In Einstein's gravity, the theory is invariant under a general coordinate transformation
\be
x^\mu \rar x'^\mu = x'^\mu (x^\mu) \, .
\ee
In conformal gravity, the theory is invariant under both a general coordinate transformation and a conformal transformation of the metric tensor $g_{\mu\nu}$
\be
g_{\mu\nu} \rar g'_{\mu\nu} = \Omega^2 g_{\mu\nu} \, ,
\ee
where $\Omega = \Omega (x)$ is a function of the spacetime point.

In Einstein's gravity, we have to clearly distinguish coordinate singularities from spacetime singularities. The former are not a physical property of the spacetime but an artifact of the coordinate system. An example is the coordinate singularity at the horizon in the Schwarzschild spacetime in Schwarzschild coordinates. The metric is ill-defined at the horizon, but the spacetime is perfectly regular there and we can perform a coordinate transformation to remove the coordinate singularity. Spacetime singularities are instead a physical property of the spacetime. An example is the singularity at the center of a Schwarzschild black hole. Curvature invariants (e.g. scalar curvature, square of the Ricci tensor, Kretschmann scalar) diverge at $r=0$ and the spacetime is geodetically incomplete there, regardless of the coordinate system.

The idea behind conformal gravity is that the spacetime singularities appearing in Einstein's gravity are actually an artifact of the gauge in conformal gravity (like coordinate singularities are an artifact of the coordinate system in Einstein's gravity) and there is always a suitable conformal transformation to remove the apparent singularity~\cite{noi1,noi2,noi3a,noi3b,noi3c}. Note that the scalar curvature, the square of the Ricci tensor, and the Kretschmann scalar are invariants in Einstein's gravity but they are not co-covariant in conformal gravity (invariant under both Weyl and general coordinate transformations), so they cannot be used to check the regularity of the spacetime. From the point of view of conformal gravity, these quantities are like the components of a tensor in Einstein's gravity that change under coordinate transformations and may diverge in some coordinate system without implying that there are spacetime singularities. Note that, for consistency, matter can only be coupled to gravity via conformally invariant terms.

The world around us is definitively not conformally invariant. If we want to consider the possibility that conformal invariance is a fundamental symmetry in Nature, we must assume it is somehow broken, and a possibility is that it is spontaneously broken. This means that Nature has to choose one of the possible vacua (exact solution of the equations of motion). The spontaneous symmetry breaking consists in replacing the selected vacuum (exact solution of the equations of motion) in the action and choose a suitable gauge (for example the unitary gauge). The field fluctuation $\varphi$ of the dilaton $\Phi$ is never physical and can be gauged away\footnote{Here we define $\Phi=\bar{\Phi}+\varphi$, where $\Phi$ is the dilaton, $\bar{\Phi}$ is the background dilaton solution, and $\varphi$ is the perturbation.}, but in the broken phase the graviton starts to propagate on the selected vacuum. The latter could be just a constant, whether the spacetime background is Minkowski (assuming vanishing cosmological constant), or a nontrivial one if we are interested in curved spacetimes. The theory in the unitary gauge is still secretly conformal invariant, but the vacuum is not~\cite{noi3a,noi3b,noi3c}. Nature should thus be able to select a vacuum which is singularity free on the base of the invariants of Einstein's gravity because these are what we can measure today in our Universe in the conformally broken phase. Note that at quantum level conformal symmetry is anomaly-free in the ultraviolet completion of Einstein's gravity as proved in~\cite{M1,M1b,M2}.

In this paper, we extend previous work and we add two more spacetimes to the list of the singular solutions of Einstein's equations that can become regular with a suitable conformal transformation $\Omega$. We consider the Janis-Newman-Winicour (JNW) metric~\cite{jnw68}, which is an exact solution of Einstein's equations in the presence of a massless scalar field, and the $\gamma$-metric~\cite{gamma1,gamma2}, which is an exact solution of Einstein's equations in vacuum. Both spacetimes have no horizon and present a naked singularity at a finite value of the radial coordinate. We show that there is an infinite family of (singular) conformal transformations that makes these spacetimes regular everywhere. Curvature invariants are finite over the whole spacetime. Massive particles cannot reach in a finite proper time the previous singular surface. Massless particles cannot do it with a finite value of their affine parameter. The spacetimes become thus geodetically complete.

The rest of the paper is organized as follows. In Section~\ref{s-review}, we briefly review the JNW and $\gamma$ solutions in Einstein's gravity. In Section~\ref{s-jnw}, we show that conformal gravity can solve the spacetime singularity in the JNW solution. In Section~\ref{s-gamma}, we repeat our study for the $\gamma$-metric and we show that conformal gravity can solve the spacetime singularity. Summary and conclusions are in Section~\ref{s-con}. Throughout the paper, we employ units in which $G_{\rm N} = c = 1$ and a metric with signature $(-+++)$.


\section{Singular spacetimes \label{s-review}}

\subsection{Janis-Newman-Winicour metric}

The JNW spacetime is an exact solution of the Einstein equations in which matter is described by a real massless scalar field $\phi$. The total action is
\be
S = \int d^4 x \, \sqrt{-g} \left[ R + g^{\mu\nu} 
\left(\partial_\mu \phi\right) \left(\partial_\nu \phi\right) \right] \, .
\ee
The static and spherically symmetric solution was found in Ref.~\cite{jnw68} and, independently, in Ref.~\cite{wyman}, and it was later shown that the two solutions were the same~\cite{v97}. The line element of the spacetime can be written as 
\be\label{e-non-rot00}
ds^2 = - A dt^2 + A^{-1} dr^2 + B \left( d\theta^2 + \sin^2\theta d\phi^2 \right) \, ,
\ee
where
\be
A &=& \left[ \frac{r - M \left(\mu - 1\right)}{r + M \left(\mu + 1\right)} \right]^{1/\mu} \, , \\
B &=& \frac{\left[ r + M \left(\mu + 1\right) \right]^{1 + 1/\mu}}{\left[ r 
- M \left(\mu - 1\right) \right]^{- 1 + 1/\mu}} \, .
\ee
The solution has thus two parameters. $M > 0$ is related to the mass of the source. $\mu \in (1,\infty)$ is the scalar charge of the source. The solution for the scalar field is
\be
\phi = \frac{\sigma}{\mu} \ln 
\left| \frac{r - M \left(\mu - 1\right)}{r + M \left(\mu + 1\right)}\right| \, ,
\ee
where $\mu$ and $\sigma$ are locked by the relation
\be\label{e-mu-sigma}
\mu = 1 + \frac{8 \pi \sigma^2}{M^2} \, .
\ee

For $\mu \neq 1$, the JNW metric has a naked singularity at the radial coordinate
\be
r_{\rm sing} = M \left(\mu - 1\right) \, .
\ee
Here the spacetime is geodetically incomplete and scalar invariants diverge
\be
&& \lim_{r \rar r_{\rm sing}} R \sim \frac{1}{\left( r - r_{\rm sing} \right)^2} \, , \nonumber\\
&& \lim_{r \rar r_{\rm sing}} R_{\mu\nu} R^{\mu\nu} 
\sim \frac{1}{\left( r - r_{\rm sing} \right)^4} \, , \nonumber\\
&& \lim_{r \rar r_{\rm sing}} R_{\mu\nu\rho\sigma} 
R^{\mu\nu\rho\sigma} \sim \frac{1}{\left( r - r_{\rm sing} \right)^4} \, .
\ee
For $\mu = 1$, we recover the Schwarzschild metric after the coordinate transformation of the radial coordinate $r' = r + 2M$. The value of the scalar charge $\mu$ indicates how much the JNW metric deviates from the Schwarzschild metric.

\subsection{$\gamma$-metric}

The $\gamma$-metric is an exact solution of the vacuum Einstein equations~\cite{gamma1,gamma2,gamma3}.
It belongs to the class of the Weyl metrics. The line element can be written as
\be\label{e-g}
\hspace{-0.5cm}
ds^2 = - F dt^2 + \frac{1}{F} \left( G dr^2 + H d\theta^2 + \Sigma \sin^2\theta d\phi^2 \right) \, ,
\ee
where
\be
F = \left( 1 - \frac{2 M}{r} \right)^\gamma \, , \quad
G = \left( \frac{\Sigma}{\Delta} \right)^{\gamma^2 - 1} \, , \quad
H = \frac{\Sigma^{\gamma^2}}{\Delta^{\gamma^2 - 1}} \, ,
\ee
$\Sigma = r^2 - 2 M r$, and $\Delta = r^2 - 2 M r + M^2 \sin^2\theta$.

As in the JNW metric, there are two parameters. $M > 0$ is related to the mass of the source. $\gamma > 0$ quantifies the deformation from spherical symmetry. For $\gamma = 1$, the spacetime is spherically symmetric and we recover the Schwarzschild solution. For $\gamma > 1$ ($\gamma < 1$), the spacetime is oblate (prolate). For $\gamma \rar 0$, we recover the flat spacetime. The total ADM mass as measured by an observer at infinity is $M_{\rm ADM} = \gamma M$.

For $\gamma \neq 1$, the spacetime has a naked singularity at the radial coordinate
\be
r_{\rm sing} = 2M \, .
\ee
Since the $\gamma$-metric is a vacuum solution, $R = R_{\mu\nu} R^{\mu\nu} = 0$. The Kretschmann scalar is non-vanishing, but its expression is quite long for a general $\gamma$. If we consider, for example, $\gamma = 2$, the Kretschmann scalar is 
\be
R_{\mu\nu\rho\sigma} R^{\mu\nu\rho\sigma} &=&
\frac{3 M^2 \left( M^2 \cos 2\theta - M^2 + 4 M r - 2 r^2 \right)^5}{r^{14} \left(r - 2 M\right)^6}
\nonumber\\
&& \cdot \left[ M^2 \left(21 M^2 - 18 M r + 4 r^2 \right) \cos 2\theta 
- 21 M^4 + 54 M^3 r - 46 M^2 r^2 + 16 M r^3 - 2 r^4 \right] \, , \nonumber\\
\ee
and is divergent at $r = r_{\rm sing}$.


\section{Resolving the singularity in the JNW metric \label{s-jnw}}

In conformal gravity, the physics is independent of the gauge, namely of the conformal factor $\Omega$. A conformal transformation in conformal gravity is on the same ground as a coordinate transformation in Einstein's gravity. In the broken phase of conformal symmetry, Nature selects one of the vacua and only general covariance remains as an explicit symmetry of the action in the conformal unitary gauge. Here we show that there is an infinite family of gauges such that the spacetime is singularity-free even using the standard tools to check the regularity of a spacetime in Einstein's gravity. Nature should be able to select one of these regular vacua.

From previous work, we expect that the conformal factor capable of solving the singularity in the JNW metric should be singular at the spacetime singularity. This is what happens even in the coordinate singularity in Einstein's gravity: the coordinate transformation to remove the coordinate singularity at the horizon in the Schwarzschild metric in Schwarzschild coordinates is singular at the event horizon. As we will show in the next subsection, the spacetime singularity of the JNW metric in Eq.~(\ref{e-non-rot00}) can be solved by the following conformal factor 
\be\label{e-om-jnw}
\Omega = \left\{ 1 + \frac{L^2}{\left[ r - M \left(\mu - 1\right) \right]^2} \right\}^{n/2} \, ,
\ee
where $n > 0$ is an integer number.

\subsection{Curvature invariants}

We have checked with the Mathematica package for tensor calculus RGTC~\cite{rgtc} the regularity of the scalar curvature, the square of the Ricci tensor, and the Kretschmann scalar. They are finite over the whole spacetime and the vanish at $r = r_{\rm sing}$ 
\be
\lim_{r \rar r_{\rm sing}} R 
= \lim_{r \rar r_{\rm sing}} R_{\mu\nu} R^{\mu\nu}
= \lim_{r \rar r_{\rm sing}} R_{\mu\nu\rho\sigma} 
R^{\mu\nu\rho\sigma} = 0 \, .
\ee
Their expressions are too long to be reported here, but they have the following form
\be
R &=& \frac{1}{
\left[ r +  r_+ \right]^2 \left[ L^2 + \left( r - r_{\rm sing} \right)^2 \right]^3}  
\times \big\{ \text{polynomial in $M$, $r$, $L$, $\mu$, $r_{\rm sing}^{1/\mu}$, and $r_+^{1/\mu}$} \big\}
\, , \nonumber\\
R_{\mu\nu} R^{\mu\nu} 
&=& \frac{1}{
\left[ r +  r_+ \right]^4 \left[ L^2 + \left( r - r_{\rm sing} \right)^2 \right]^6}  
\times \big\{ \text{polynomial in $M$, $r$, $L$, $\mu$, $r_{\rm sing}^{1/\mu}$, and $r_+^{1/\mu}$} \big\} \, , \nonumber\\
R_{\mu\nu\rho\sigma} R^{\mu\nu\rho\sigma}
&=& \frac{1}{
\left[ r +  r_+ \right]^4 \left[ L^2 + \left( r - r_{\rm sing} \right)^2 \right]^6}  
\times \big\{ \text{polynomial in $M$, $r$, $L$, $\mu$, $r_{\rm sing}^{1/\mu}$, and $r_+^{1/\mu}$} \big\} \, . \nonumber\\
\ee
where $r_+ = M \left(\mu + 1\right)$. We can thus see that divergent terms in the JNW metric with the form
\be
\frac{1}{\left( r - r_{\rm sing} \right)^2}
\ee
are now replaced by
\be
\frac{1}{L^2 + \left( r - r_{\rm sing} \right)^2} \, ,
\ee
which reduces to $1/L^2$ for $r \rar r_{\rm sing}$ and do not diverge.

\subsection{Time-like geodesics}

Let us now check if the spacetime is geodetically complete at the singular surface $r = r_{\rm sing}$ of the JNW metric. For massive particles, we have $g_{\mu\nu} \dot{x}^\mu \dot{x}^\nu = - 1$, where the dot $\dot{}$ indicates the derivative with respect to the proper time $\tau$. For simplicity, we consider purely radial trajectories ($\dot{\theta} = \dot{\phi} = 0$) and therefore our master equation is
\be\label{eq-ds2}
g_{tt} \dot{t}^2 + g_{rr} \dot{r}^2 = - 1 \, .
\ee
Since the metric coefficients are independent of the temporal coordinate $t$, we have the conservation of the particle energy $E$
\be\label{eq-ds2a}
p_t = g_{tt} \dot{t} = - E \, .
\ee
If we employ Eq.~(\ref{eq-ds2a}) into Eq.~(\ref{eq-ds2}), we get
\be
\dot{r}^2 = - \frac{E^2 + g_{tt}}{g_{tt} g_{rr}} \, .
\ee
In our case of the conformally modified JNW metric, this equation reads
\be\label{e-jnw-timelike}
\dot{r}^2 = \frac{E^2 - \Omega^2 A}{\Omega^4} \, .
\ee
Integrating by part, we find that a particle moving from an initial radial coordinate $r_{\rm in}$ towards $r_{\rm sing}$ takes an infinite proper time to reach a certain radius $r_* \ge r_{\rm sing}$   
\be
\tau = \int_{r_*}^{r_{\rm in}} \frac{\Omega^2 dr}{\sqrt{E^2 - \Omega^2 A}}
\rar + \infty \, .
\ee
$r_*$ is either the radial coordinate at which the denominator of the integrand vanishes or $r_{\rm sing}$ if the denominator of the integrand does not vanish. The left panel in Fig.~\ref{f-jnw} shows the numerical integration of this expression for $E=M=L=1$, $n=\mu=2$, and $r_{\rm in}=10$.

\subsection{Light-like geodesics}

\begin{figure*}[t]
\begin{center}
\includegraphics[type=pdf,ext=.pdf,read=.pdf,width=7.0cm]{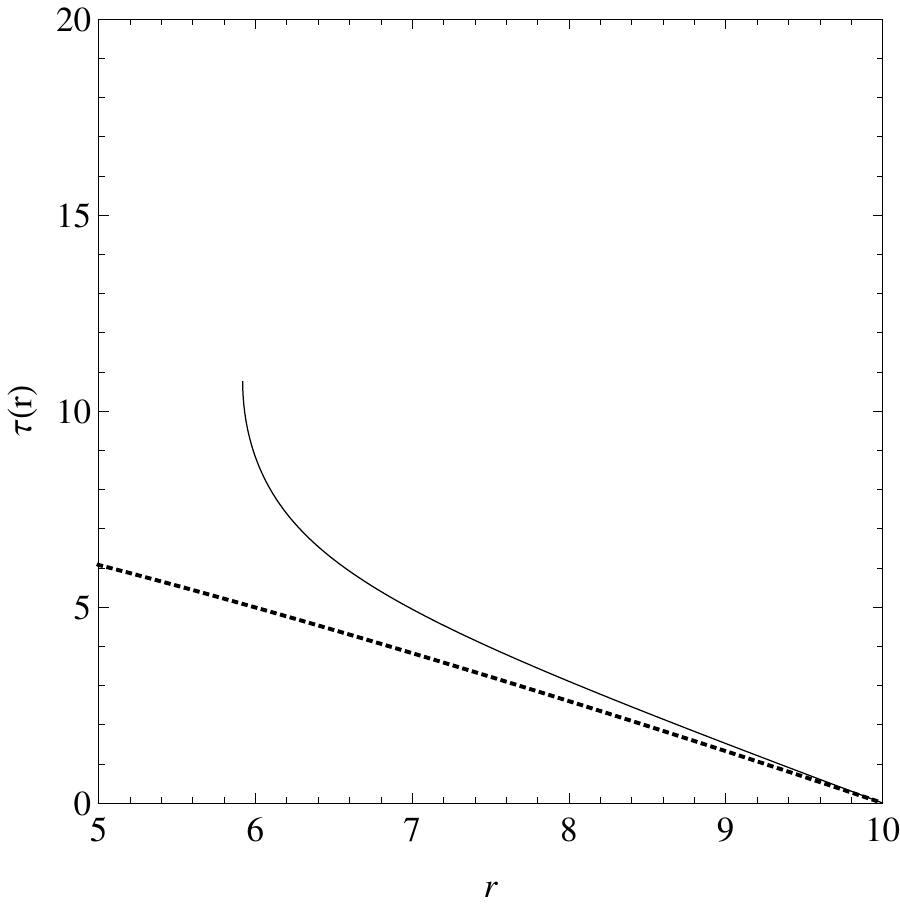}
\hspace{0.8cm}
\includegraphics[type=pdf,ext=.pdf,read=.pdf,width=7.0cm]{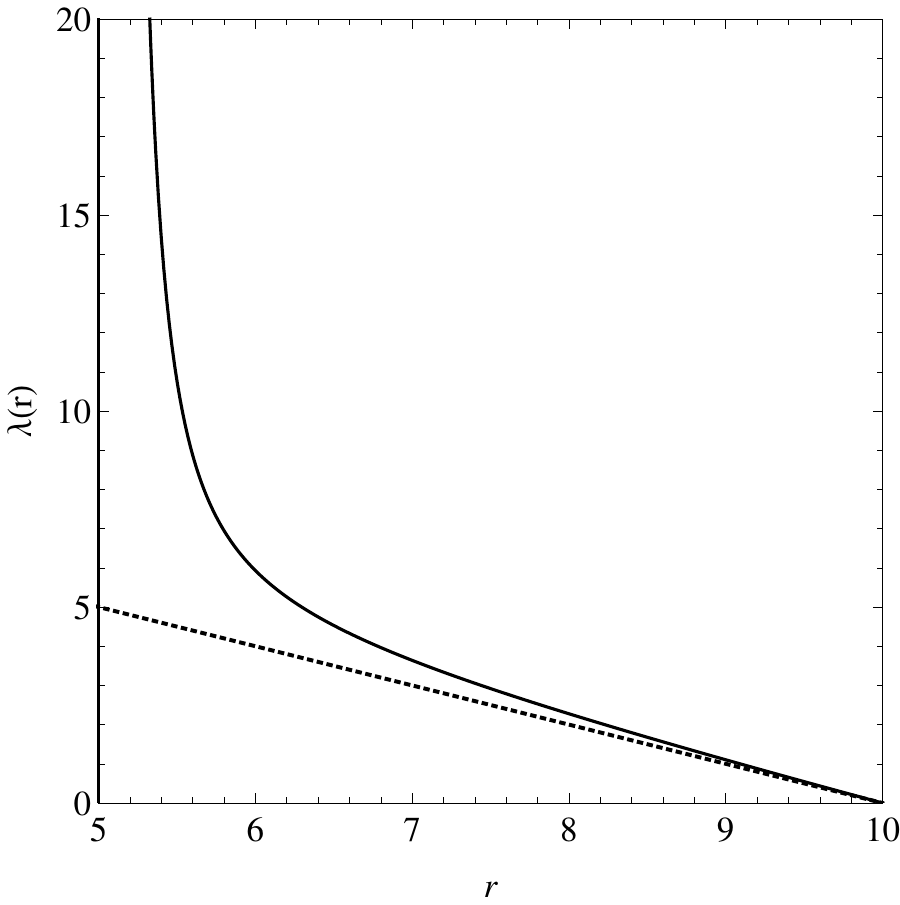}
\end{center}
\vspace{-0.5cm}
\caption{Left panel: proper time $\tau$ as a function of the radial coordinate $r$ of a massive particle with vanishing angular momentum moving to smaller radii from the initial coordinate $r = r_{\rm in}$. The solid line corresponds to the proper time in the rescaled metric and the particle cannot reach the surface $r = r_{\rm sing}$. The dashed line corresponds to the standard JNW metric in Einstein's gravity. Right panel: as in the left panel for the affine parameter $\lambda$ of a massless particle. In these plots, we assume $E=M=L=1$, $n = \mu = 2$, and $r_{\rm in}=10$. \label{f-jnw}}
\end{figure*}

The generalization to massless particles is straightforward. The counterpart of Eq.~(\ref{eq-ds2}) is
\be\label{eq-massless}
g_{tt} \dot{t}^2 + g_{rr} \dot{r}^2 = 0 \, ,
\ee
where now the dot $\dot{}$ indicates the derivative with respect to an affine parameter $\lambda$. Now we find
\be
\dot{r}^2 = - \frac{E^2}{g_{tt} g_{rr}} \, ,
\ee
and, when we integrate by parts, we get
\be
\lambda = \int_{r_{\rm sing}}^{r_{\rm in}} \frac{\Omega^2 dr}{E}
\rar + \infty \, ,
\ee
for any integer value of $n > 0$. No massless particle can reach the surface $r = r_{\rm sing}$ with a finite value of the affine parameter $\lambda$. The right panel in Fig.~\ref{f-jnw} shows the numerical integration of this formula for $E=M=L=1$, $n=\mu=2$, and $r_{\rm in}=10$.

\section{Resolving the singularity in the $\gamma$-metric \label{s-gamma}}

With the same spirit, we can look for a suitable $\Omega$ to resolve the spacetime singularity in the $\gamma$-metric~(\ref{e-g}). For the sake of simplicity, we discuss the case $\gamma=2$, but we expect that a generalization is possible for any real value of $\gamma$. The form factor capable of resolving the spacetime singularity is the counterpart of that in Eq.~(\ref{e-om-jnw}) for the case $r_{\rm sing} = 2M$
\be
\Omega = \left[ 1 + \frac{L^2}{\left( r - 2M \right)^2} \right]^{n} \, ,
\ee
where $n > 0$ is an integer number. Note that the exponent here is $n$, while it was $n/2$ in Eq.~(\ref{e-om-jnw}). If the exponent were $n/2$ here, the curvature invariants would still diverge with $n=1$.

\subsection{Curvature invariants}

We proceed as in the case of the JNW metric and we use the Mathematica package for tensor calculus RGTC. For $n=1$, the expressions for the scalar curvature and the square of the Ricci tensor are compact and read
\be
R &=& -\frac{3 L^2 \left(r^2-4 M^2\right) \left(M^2-4 M r+2
   r^2 - M^2 \cos 2 \theta \right)^3}{2 r^6 \left[L^2+(r-2 M)^2\right]^3} \, , \nonumber\\
R_{\mu\nu} R^{\mu\nu} &=& \frac{L^4 (r-2 M)^2 
\left(M^2-4 M r+2 r^2 - M^2 \cos 2 \theta \right)^5 }{4 r^{12} \left[L^2+(r-2
   M)^2\right]^8} \nonumber\\
   && \times \big\{ L^4
   \left(82 M^4-162 M^3 r+55 M^2 r^2-4 M r^3+6 r^4\right) 
   \nonumber\\ && \qquad +4 L^2 (r-2 M)^2 \left(41
   M^4-92 M^3 r+53 M^2 r^2-10 M r^3+2 r^4\right)
   \nonumber\\ && \qquad -M^2 \cos 2 \theta \big[ L^4
   \left(82 M^2-26 M r+3 r^2\right)+4 L^2 (r-2 M)^2 \left(41 M^2-24 M r+7
   r^2\right) \nonumber\\ && \hspace{3.0cm}
   +(r-2 M)^4 \left(82 M^2-70 M r+37 r^2\right)\big] 
   \nonumber\\ && \qquad +(r-2 M)^4 \left(82
   M^4-206 M^3 r+169 M^2 r^2-84 M r^3+26 r^4\right)\big\} \, .
 \ee
The expression for the Kretschmann scalar is still long, but has the form
\be
R_{\mu\nu\rho\sigma} R^{\mu\nu\rho\sigma} &=& 
\frac{(r-2 M)^2}{4 r^{14} \left[L^2+(r-2 M)^2\right]^8} \times \big\{ \text{polynomial in $M$, $r$, and $L$} \big\}\, .
\ee
If $L \neq 0$, these scalars do not diverge for $r \rar 2M$.

\subsection{Time-like geodesics}

Again, we proceed as in the case of the JNW metric. The counterpart of Eq.~(\ref{e-jnw-timelike}) is
\be
\dot{r}^2 = \frac{E^2 - \Omega^2 F}{\Omega^4 G} \, .
\ee
For a massive particle moving from an initial radial coordinate $r_{\rm in}$ towards $r_{\rm sing}$, the proper time is given by 
\be
\tau = \int_{r_*}^{r_{\rm in}} \frac{\Omega^2 G^{1/2} dr}{\sqrt{E^2 - \Omega^2 F}}
\rar + \infty \, ,
\ee
and diverges at $r_* \ge r_{\rm sing}$, as it was in the JNW spacetime. The left panel in Fig.~\ref{f-gamma} shows the numerical integration of this expression for $E=M=L=1$, $n=2$, and $r_{\rm in}=10$.

\subsection{Light-like geodesics}

\begin{figure}[t]
\begin{center}
\includegraphics[type=pdf,ext=.pdf,read=.pdf,width=7.0cm]{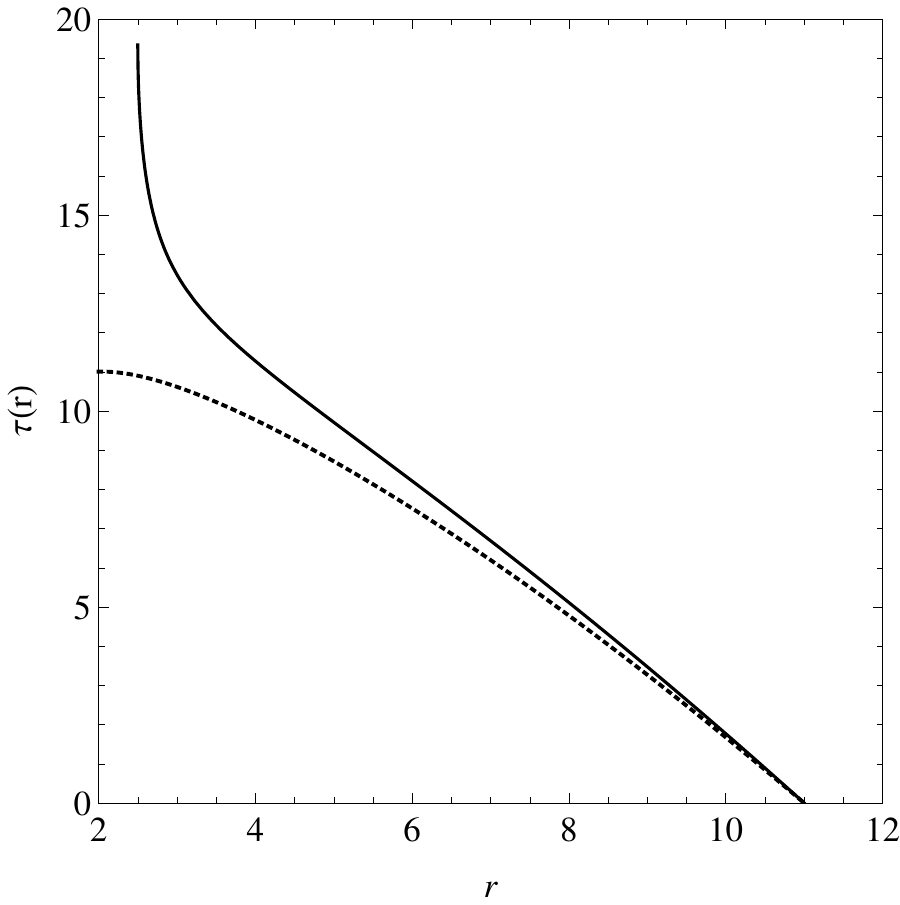}
\hspace{0.8cm}
\includegraphics[type=pdf,ext=.pdf,read=.pdf,width=7.0cm]{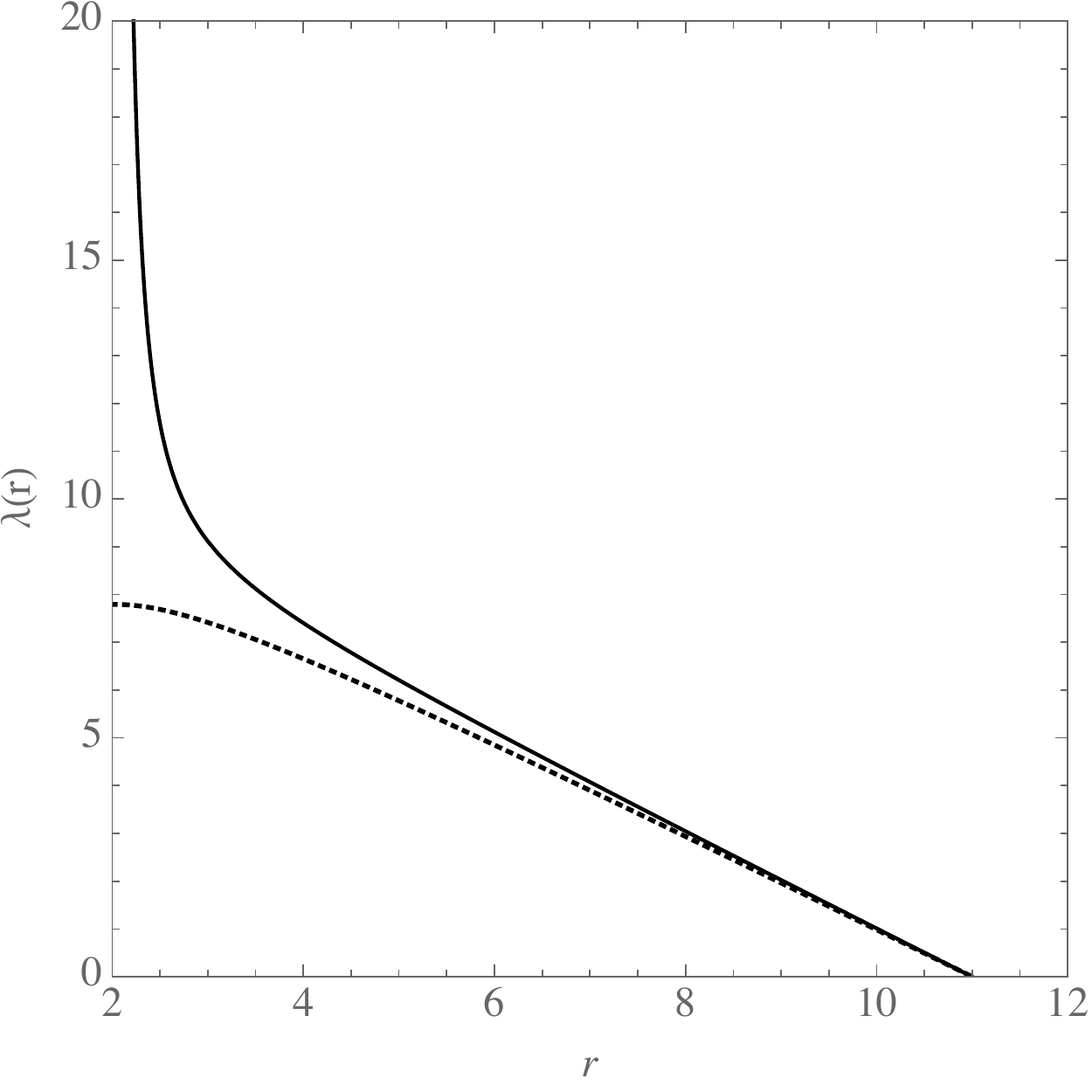}
\end{center}
\vspace{-0.5cm}
\caption{As in Fig.~\ref{f-jnw} for the $\gamma$-metric. The left panel shows the case for a massive particle, the right panel that for a massless one. In these plots, we assume $E=M=L=n=1$, $\gamma = 2$, and $r_{\rm in}=11$.\label{f-gamma}}
\end{figure}

We employ Eq.~(\ref{eq-massless}), where now $g_{tt} g_{rr} = - \Omega^4 G$. When we integrate by parts, we get
\be
\lambda = \int_{r_{\rm sing}}^{r_{\rm in}} \frac{\Omega^2 G^{1/2} dr}{E}
\rar + \infty \, .
\ee
No massless particle can reach the surface $r = r_{\rm sing}$ with a finite value of the affine parameter $\lambda$. The right panel in Fig.~\ref{f-gamma} shows the numerical integration of this expression for $E=M=L=1$, $n=2$, and $r_{\rm in}=10$.

\section{Concluding remarks \label{s-con}}

Einstein's gravity is plagued by solutions with spacetime singularities, where predictability is lost and standard physics breaks down. Despite significant efforts in the past sixty years, it is extremely difficult to get a gravity theory completely free from singular solutions. Conformal gravity seems to be able to solve the singularity problem in an elegant way by enlarging the symmetry group. Previous work has shown how conformal gravity can solve the spacetime singularities in the Friedmann-Robertson-Walker~\cite{confgrav2}, Schwarzschild and Kerr metrics~\cite{noi1}. In the present paper, we have added two more spacetimes to this list, the JNW metric~\cite{jnw68} and the $\gamma$-metric~\cite{gamma1,gamma2}.

The JNW metric is an exact solution of Einstein's gravity in the presence of a minimally coupled real massless scalar field. The $\gamma$-metric is instead an exact solution of Einstein's gravity in vacuum. Both spacetimes have no event horizon and possess a naked singular surface at a finite value of the radial coordinate $r$. The spacetime is singular because it is geodetically incomplete and curvature invariants diverge at this singular surface.

In this paper, we have shown that conformal invariance can solve the spacetime singularities in the JNW and $\gamma$-metrics. Now curvature scalars are always finite. The spacetimes are geodetically complete and the previous singular surfaces are now ``unattainable spacetime regions''. No massive particle can reach the radial coordinate $r = r_{\rm sing}$. No massless particle can do the same within a finite value of its affine parameter. We emphasize that our results support the conjecture according to which conformal gravity in the spontaneously broken phase can fix the singularity problem.


\begin{acknowledgments}
This work was supported by the National Natural Science Foundation of China (Grant No.~U1531117) and Fudan University (Grant No.~IDH1512060). H.C. also acknowledges support from the China Scholarship Council (CSC), grant No.~2017GXZ019020. C.A.B.G. also acknowledges support from the China Scholarship Council (CSC), grant No.~2017GXZ019022. C.B. also acknowledges the support from the Alexander von Humboldt Foundation.
\end{acknowledgments}


\end{document}